\documentclass{optica-article}

\journal{opticajournal} 

\articletype{Research Article}
\usepackage{graphicx}%
\usepackage{caption}
\usepackage{subcaption}
\usepackage{lineno}
\usepackage{xcolor}

\newtheorem{theorem}{Theorem}[section]

\begin{document}

\title{Estimating the time-evolving refractivity of a turbulent medium using optical beam measurements: a data assimilation approach}

\author{Anjali Nair, \authormark{1,*} Qin Li,\authormark{2} and  Samuel N. Stechmann\authormark{2,3}}

\address{\authormark{1}Department of Statistics and CCAM, University of Chicago, Chicago, IL 60637 USA\\
\authormark{2}Department of Mathematics, University of Wisconsin - Madison, Madison, WI 53706 USA\\
\authormark{3}Department of Atmospheric and Oceanic Sciences, University of Wisconsin - Madison, Madison, WI 53706 USA}

\email{\authormark{*}anjalinair@uchicago.edu} 


\begin{abstract*}   In applications such as free-space optical communication, a signal is often recovered after propagation through a turbulent medium. In this setting, it is common to assume that limited information is known about the turbulent medium, such as a space- and time-averaged statistic (e.g., root-mean-square), but without information about the state of the spatial variations.  It could be helpful to gain more information if the state of the turbulent medium can be characterized with the spatial variations and evolution in time described. Here, we propose to investigate the use of data assimilation techniques for this purpose. A computational setting is used with the paraxial wave equation, and the extended Kalman filter is used to conduct data assimilation using intensity measurements. To reduce computational cost, the evolution of the turbulent medium is modeled as a stochastic process. Following some past studies, the process has only a small number of Fourier wavelengths for spatial variations. The results show that the spatial and temporal variations of the medium are recovered accurately in many cases. 
In some time windows in some cases, the error is larger
for the recovery.
Finally we discuss the potential use of the 
spatial variation information for aiding the recovery
of the transmitted signal or beam source.
\end{abstract*}

\section{Introduction}

It is well known that turbulence can have a degrading effect in the performance of optical communications~\cite{andrews2005laser,gbur2002spreading, feizulin1967broadening, baykal1983scintillations, korotkova2004model, schulz2005optimal, avramov2014polarization, rabinovich2015free, Rabinovich:16, Beason:20, borcea2020multimode, bos2022wave, li2022computation, nair2023scintillation}. This includes unwanted effects like reduced signal intensities, scintillation and beam wander which in turn affect the recovery of the true signal.

\begin{figure}[h]
           \hspace{0.5cm}
\begin{subfigure}[b]{0.3\textwidth}  
       \includegraphics[width=4cm]{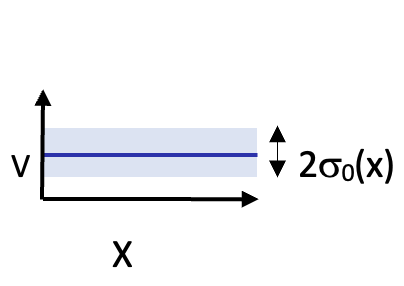}
    \caption{ }
    \label{fig:data_assim1}
    \end{subfigure}
    \begin{subfigure}[b]{0.3\textwidth}    
  \includegraphics[width=4cm]{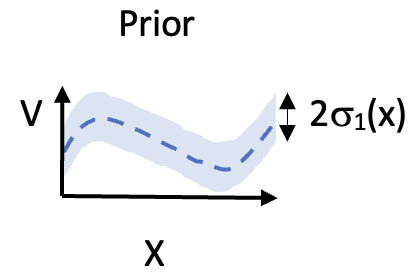}
    \caption{ }
    \label{fig:data_assim2b}
    \end{subfigure}
        \begin{subfigure}[b]{0.3\textwidth}    
  \includegraphics[width=4cm]{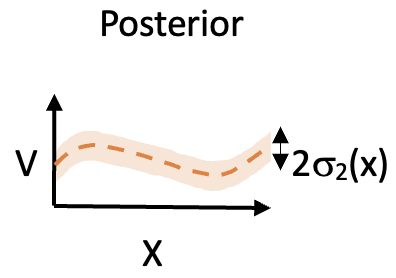}
    \caption{ }
    \label{fig:data_assim2c}
    \end{subfigure}
      \caption{Schematic illustration of estimates of refractivity, with uncertainty quantification. (a) A traditional estimate of the mean and standard deviation of the refractivity (e.g., mean refractivity and $C_n^2$ coefficient, respectively), which lacks knowledge of spatial variations. (b-c) An estimate that includes the spatial variations of the {mean} refractivity, $V(x)$, along with spatial variations of the uncertainty, $\sigma(x)$. Furthermore, the aim is to reduce the uncertainty from $\sigma_1(x)$ to $\sigma_2(x)$ as more measurements are accumulated over time, as indicated in moving from the prior to posterior estimates.}
    \end{figure}

In this setting, it is common to assume that we have limited knowledge about the turbulent medium, such as a limited statistic (e.g., root-mean-square, $C_n^2$ parameter), but without information about the state of the spatial variations, as illustrated in Fig.~\ref{fig:data_assim1}. Optical turbulence is typically modelled by randomness in the refractive index of the medium.
Any knowledge about the current state of the refractive index can be potentially very useful in providing a more accurate description of how the medium affects signal performance. It can also aid in signal recovery.

Estimating the current state of a medium is a rather challenging problem, as the medium often fluctuates substantially in both space and time. It would be desirable to have estimates of the space and time variations of the medium, as illustrated in Fig.~\ref{fig:data_assim2b} and Fig.~\ref{fig:data_assim2c}. Generally speaking, in many areas of science and engineering, one typically learns the state of a medium by probing it with known signals and measuring the corresponding outputs at the receiver. Some related studies on estimating refractivity using radar can be found in~\cite{zhao2011theoretical, gilles2019subspace, karabacs2021variational}. Reconstructing optical turbulence using measurements is still not fully understood, in large part due to the complex interplay between the observables and the fluctuations in the medium. 

Here we investigate the possibility of estimating the space and
time variations of the turbulent medium, using intensity measurements at the receiver, received from an optical beam source. In recent decades, much progress has been made on 
similar science and engineering problems with methodologies that have been developed in different areas~\cite{lorenc1986analysis, oliver2008inverse, iglesias2013evaluation, karcher2018reduced}. The common theme of these problems is how to estimate the current state of a system. These methodologies go under different names, such as data assimilation~\cite{zupanski1997general, hunt2007efficient,evensen2009data,law2012evaluating,chen2019multi,chen2022superfloe,sanz2023inverse}, 
state estimation~\cite{stengel1994optimal,simon2006optimal,majda2018model}, 
parameter estimation~\cite{beck1977parameter, chen2014mcmc}, filtering~\cite{kalman1960new,jazwinski2007stochastic,tippett2003ensemble,anderson2012optimal, chorin2010implicit}, smoothing~\cite{evensen2000ensemble}, and inverse problems~\cite{stuart2010inverse, kaipio2006statistical,martin2012stochastic,ding2021ensemble,gamba2022reconstructing,Chen2018,du2023inverse}. 
These methods are potentially applicable in the present context as well.
Data assimilation techniques have been explored in the past to obtain refractivity estimates using not optical signals but with radar~\cite{yardim2008tracking}. However, this is still underexplored for applications using optical signals and we provide some first-stage results in this pursuit.


We propose to investigate the use of data assimilation methods to estimate the refractive index of the turbulent medium. One prominent data assimilation technique is the Kalman filter. The basic idea of data assimilation and the Kalman filter is to combine two sources of information (typically a prior estimate from a model, and an estimate from observations or measurements) in order to merge their information and find a better estimate. A schematic illustration is shown in Fig.~\ref{fig:data_assim2b} and Fig.~\ref{fig:data_assim2c}, where the mean state of the medium is updated, and the uncertainty changes from a larger uncertainty value $\sigma_1(x)$ to a smaller uncertainty $\sigma_2(x)$ due to the new information gained from a new measurement. The terminology of ``assimilation'' arises from the setting where a time-evolving model is taking in or assimilating observational data in order to sequentially adjust its state. 

In the present paper we investigate the problem in a computational setting, where the turbulent medium and optical signal are modeled numerically.
A time evolving medium can in principle be modelled using fundamental conservation laws like the Navier--Stokes equations; however, owing to the fine scale dynamics, this is not tractable in practice. It is then common to model the time evolution of the state of the medium as a random process. This can be done by describing the spatial variations in the random medium through a number of Fourier modes, with each mode modelled by an independent stochastic process~\cite{delsole2004stochastic, harlim2008filtering, majda2012filtering, qi2017low, branicki2018accuracy, li2020predictability} in time.

One challenge here is that the measurements arise from a beam that propagates through only a small fraction of the overall spatial volume of a medium. Hence, only a small fraction of the domain is probed by each beam measurement.
One might wonder whether much information can be recovered about a turbulent medium in such a situation.
To help with this situation of sparse data in space, one can leverage data at many different instances in time. Such a situation arises in other settings as well, such as sparse measurements of the atmosphere in weather forecasting; and in that case, data assimilation techniques are nevertheless able to provide a reasonably accurate recovery of the state of the atmosphere. Here, in the setting of optical beam measurements, we investigate a similar question: whether useful information can be gained from an accumulation of {sparse} beam measurements at a sequence of different times.  

The paper is organized as follows. In Section~\ref{sec:math_setup}, we outline the mathematical setup of the problem. In particular, we describe the paraxial wave equation (PWE), which provides the forward map from the refractive index as input to the measurements as output, and also the associated inverse problem. In Section~\ref{sec:data_assimilation}, we describe the data assimilation framework. We first review the general setup in Section~\ref{subsec:Filtering_strategies} and then illustrate how this can be applied to the setting of the PWE in Section~\ref{subsec:data_assimi_PWE}. The results of numerical studies are presented in Section~\ref{sec:numerical_exp}.

\section{Mathematical setup: forward and inverse problems for the paraxial wave equation}\label{sec:math_setup}
We consider the setting of an optical beam that propagates through a turbulent medium.
If the medium fluctuates at a slower rate compared to the time it takes the optical signal to reach the receiver, we can assume that the medium is fixed and deterministic during the entire propagation interval. In this setting, we use the PWE to describe the spatial evolution of the signal:
\begin{equation}\label{eqn:PWE}
\begin{aligned}
    \nabla_{X}^2A+2ik\partial_zA+k^2VA&=0,\\
     A(X,0)&=\phi(X)\,.
\end{aligned}
\end{equation}
Here, $A$ is the complex-valued amplitude of the signal, $z$ is the direction of propagation of the signal, and $X=(x,y)\in\mathbb{R}^2$ are the coordinates transverse to the direction of propagation. Also, $V=V(X,z)=n_r^2(X,z)-1$ where $n_r$ is the unknown refractive index of the medium. The initial value $\phi$ is a known optical source, and we measure the intensity of the signal at location $z=Z$. 

Measurements are typically taken by a weighted average in a receiver region. For example, one can use a window function $\psi$ such that
\begin{equation}\label{eqn:psi}
    \begin{aligned}
        \psi(X)&=\begin{cases}
            1,\quad X\in\mathcal{R},\\
            0,\quad X\notin\mathcal{R}\,,
        \end{cases}
    \end{aligned}
\end{equation}
where $\mathcal{R}$ denotes the receiver region. So we have measurements of the form
\begin{equation}
    d=\int\limits_{X\in\mathbb{R}^2}|A(X,z=Z)|^2\psi(X)\mathrm{d}X\,.
\end{equation}
In practice, we can replace the discontinuous window function in~\eqref{eqn:psi} by a smooth tapered window. This allows us to define a measurement map $\mathcal{M}(V):(\phi,\psi)\to d$ such that
\begin{equation}\label{eqn:forward_operator}
    \mathcal{M}(V):(\phi,\psi)\to\int\limits_{X\in\mathbb{R}^2}|A_{V,\phi}(X,z=Z)|^2\psi(X)\mathrm{d}X=d\,,
\end{equation}
where the subscript in $A_{V,\phi}$ denotes the dependence of the optical signal $A$ on the refractive index $V$ and source $\phi$. 
Suppose we have $J_1$ different sources, denoted $\phi_{j_1}$, for $j_1=1,\cdots,J_1$, and $J_2$ different receiver windows, denoted $\psi_{j_2}$, for $j_2=1,\cdots,J_2$, as in, for instance, a multi-aperture receiver with a spatial array of $J_2$ receiver apertures. Then we have $J_1\times J_2$ different data configurations:
\begin{equation}\label{eqn:d_ij}
    d_{j_1,j_2}=\int\limits_{X\in\mathbb{R}^2}|A_{V,\phi_{j_1}}(X,z=Z)|^2\psi_{j_2}(X)\mathrm{d}X, \quad  j_1=1,\cdots,J_1, \quad j_2=1,\cdots,J_2,
\end{equation}
where $A_{V,\phi_{j_1}}$ denotes the signal due to source $\phi_{j_1}$ propagating through a medium characterized by $V$. We can also denote the corresponding measurement maps by $\mathcal{M}_{j_1,j_2}(V):(\phi_{j_1},\psi_{j_2})\to d_{j_1,j_2}$, for $j_1=1,\cdots,J_1$ and $j_2=1,\cdots,J_2$.

As in the setting of inverse problems
\cite{stuart2010inverse, kaipio2006statistical,martin2012stochastic,ding2021ensemble,gamba2022reconstructing,Chen2018,du2023inverse}, 
one can aim to reconstruct $V$ using the data $\{d_{j_1,j_2}\}$, for $j_1=1,\cdots,J_1$ and $j_2=1,\cdots,J_2$. This is commonly done in a least squares minimization setup with a cost function $\mathcal{L}$:
\begin{equation}\label{eqn:cost_fn}
    \min_V\mathcal{L}(V),\quad\text{with }\quad  \mathcal{L}(V)=\frac{1}{2J_1J_2}\sum_{j_1,j_2}L_{j_1,j_2}^2\,,
\end{equation}
where 
\begin{equation}
    L_{j_1,j_2}=\mathcal{M}_{j_1,j_2}(V)-d_{j_1,j_2}
\end{equation}
is a measure of mismatch between the predicted measurements and true data.  In practice, most data is polluted by measurement noise. Here we assume that the noisy data is given by:
\begin{equation}
  d_{j_1,j_2}= \mathcal{M}_{j_1,j_2}(V) + \sigma_{j_1,j_2}\,,
\end{equation}
where $\sigma_{j_1,j_2}$ is a random variable drawn from $\mathcal{N}(0,\Sigma_0)$, the Gaussian (or normal) distribution with mean $0$ and covariance $\Sigma_0$.

Suppose we also have some prior knowledge that the true state of the medium is close to $V_0$, then we can reformulate the loss function to incorporate the prior information:
\begin{equation}\label{eqn:cost_Bayesian_optimization}
    \mathcal{L}(V)= \frac{1}{2J_1J_2}\sum_{j_1,j_2}\|\mathcal{M}_{j_1,j_2}(V)-d_{j_1,j_2}\|_{\Sigma_0^{-1}}^2 + \frac{1}{2}\|V-\bar{V_0}\|^2_{\Sigma^{-1}}\,,
\end{equation}
where $\bar{V}_0$ denotes the prior mean and $\Sigma$ is the prior covariance. Here, for any matrix $\Sigma$, the notation $\|\cdot\|_\Sigma$ stands for the weighted norm:
\begin{equation}
    \|f\|_\Sigma=\sqrt{f^\ast\Sigma f}\,,
\end{equation}
with $f^\ast$ being the complex conjugate of $f$. The two terms in (\ref{eqn:cost_Bayesian_optimization}) represent the two sources of information about the medium: the measurements and the prior.
Furthermore, these two terms are weighted according to their uncertainties, as quantified by their covariance matrices, $\Sigma_0$ and $\Sigma$, so that more weight is assigned when the uncertainty is lower. Hence, the measurement or prior is trusted more when its uncertainty is lower.

While the setup above is for a single instant of time for the medium $V(X,z)$, it can also be used as the foundation for a setting with a time-evolving medium $V(X,z,t)$. For instance, the $V(X,z)$ above could represent the present time, and the prior $V_0$ in (\ref{eqn:cost_Bayesian_optimization}) could represent knowledge of the medium obtained from earlier times. Such a time-evolving setting is described in the next section.



\section{Estimating the medium using data assimilation}
\label{sec:data_assimilation}

In this section, we describe how data assimilation can be used to estimate the state of a turbulent medium, using a sequence of beam measurements from different times.
First the time evolution of the medium is described in 
subsection~\ref{subsec:time-evol-medium},
and then the data assimilation approach is described
in general form in subsection~\ref{subsec:Filtering_strategies}
and specifically for the PWE in subsection~\ref{subsec:data_assimi_PWE}.

\subsection{Time evolution of the turbulent medium}
\label{subsec:time-evol-medium}

The state of the medium $V$ often evolves in time following complex conservation laws arising from fluid dynamics, such as the Navier--Stokes equations. While it is possible to apply data assimilation to simulations of the Navier--Stokes equations, it is very computationally expensive. Consequently, it is common to seek simpler models for the evolution of a turbulent fluid that are computationally cheaper while still maintaining as much realism as possible. In this direction, many past studies have shown that random processes can be used as a surrogate model or replacement for a complex turbulent dynamical system, and in data assimilation this strategy can yield promising results~\cite{delsole2004stochastic, harlim2008filtering, majda2012filtering, qi2017low, branicki2018accuracy, li2020predictability}. 

Here we will follow such a strategy of using a stochastic process as a surrogate model for the complex dynamics of the turbulent medium.
Typically, the unknown parameter is expanded in terms of a few Fourier modes, with each mode obeying an independent random process. In this spirit, and following past work \cite{delsole2004stochastic, harlim2008filtering, majda2012filtering, qi2017low, branicki2018accuracy, li2020predictability}, we use an Ornstein--Uhlenbeck (OU) process to describe the evolution of $V$ in time. Suppose $V$ consists of a linear combination of Fourier modes of the form:
\begin{equation}
    V(X,z,t)=\sum\limits_{k_X,k_z} \hat{V}_{k_X,k_z}(t) e^{i (k_X\cdot X + k_z\cdot z)}\,,
\end{equation}
where each of the coefficients $\hat{V}_{k_X,k_z}(t)$ obeys an OU process:
\begin{equation}\label{eqn:OU_process}
    \dot{\hat{V}}_{k_X,k_z}=-\gamma_{k_X,k_z}\hat{V}_{k_X,k_z} + \sigma_{k_X,k_z} \dot{W}\,.
\end{equation}
Here, $\dot{W}(t)$ is white noise associated with a standard Brownian motion $W(t)$. 
The parameter $\sigma_{k_X,k_z}>0$ is a measure of the strength of fluctuations of this random process. Note that the OU process is a continuous-time model, and it is analogous to an autoregressive model, which is a discrete-time model.

A key time scale of the turbulent medium is its decorrelation time scale, which is given by ${\gamma^{-1}_{k_X,k_z}}$, and which we assume to be much larger than the time it takes the optical signal to arrive at the receiver. This means that for each measurement snapshot at the detector, the state of the medium $V$ can be viewed to be frozen in time. Nevertheless, data assimilation allows for information to be accumulated from measurements at different times, as described next.

\subsection{Data assimilation strategy in general form}
\label{subsec:Filtering_strategies}

The two basic ingredients for data assimilation
(as described briefly in the Introduction section)
are two sources of information about a system:
a time-evolving model and measurements or observations.
These two sources of information are both imperfect or uncertain,
and the aim is to combine them in order to obtain a new estimate
with reduced uncertainty.

A variety of methods for data assimilation/filtering have been introduced
and have different advantages and disadvantages in different
scenarios. 
For instance, the original Kalman filter algorithm~\cite{kalman1960new} was formulated for the case when the model and measurements are both linear, and when the uncertainties or errors obey a Gaussian (i.e., normal) distribution. In this Gaussian case, one favorable aspect is that one can uniquely characterize all distributions using their means and covariances. However, this linear, Gaussian case is often not the case that arises in practice. To address settings that are non-linear and non-Gaussian, many extensions of the original Kalman filter algorithm and other data assimilation algorithms have been developed
\cite{lorenc1986analysis, oliver2008inverse, iglesias2013evaluation, karcher2018reduced,zupanski1997general, hunt2007efficient,evensen2009data,law2012evaluating,chen2019multi,chen2022superfloe,sanz2023inverse,kalman1960new,jazwinski2007stochastic,tippett2003ensemble,anderson2012optimal, chorin2010implicit}.

In the present paper, we investigate a data assimilation method called the extended Kalman filter \cite{majda2012filtering,sanz2023inverse}. It is chosen for investigation here for its appropriateness for the present setting, its past success on many other problems, and the simplicity of its formulation.
The extended Kalman filter provides an extension of the (linear) Kalman filter to a nonlinear setting, where the basic idea is to linearize the nonlinearities. A new linearization is typically performed at each time step, by linearizing about the current state of the system, in order to provide an accurate linearization.
One advantage of the extended Kalman filter is that, with its linearization, it can be solved with closed-form solutions.

In what follows, we describe the (linear) Kalman filter algorithm in its general form, followed by the extended Kalman filter.

Let $V_n$ be the state of the unknown parameter at time $t_n$. Suppose $V_{n+1}$ follows the linear difference equation:
\begin{equation}\label{eqn:V_model_eqn}
    V_{n+1}=F_{n+1}V_{n}+w_{n+1}\,,
\end{equation}
where $F_{n+1}$ is a linear operator and $w_{n+1}\sim\mathcal{N}(0,\Sigma_{n+1})$ is a Gaussian distributed random variable that incorporates the model error. Given that the mean and covariance of the state at time $t_n$ are $\bar{V}_{n|n}$ and $R_{n|n}$ respectively, the model equation Eq.~\eqref{eqn:V_model_eqn} provides a prediction of the mean and covariance of the state at time $t_{n+1}$ given by
\begin{equation}
    \begin{aligned}\label{eqn:prior_update}
        \bar{V}_{n+1|n}&=F_{n+1}\bar{V}_{n|n},\\
        R_{n+1|n}&=F_{n+1}R_{n|n}F_{n+1}^\top +\Sigma_{n+1}\,.
    \end{aligned}
\end{equation}
The quantities $\bar{V}_{n+1|n}$ and $R_{n+1|n}$ are called the prior mean and prior covariance, respectively, predicted by the model~\eqref{eqn:V_model_eqn}. 

Now suppose that, at each time step, we bring in additional information in the form of observations or measurements as
\begin{equation}\label{eqn:data_KF}
    d_{n+1}=G_{n+1}V_{n+1}+\sigma_{n+1}\,,
\end{equation}
where $d_{n+1}$ is a vector of observations, $G_{n+1}$ is a linear operator and $\sigma_{n+1}\sim\mathcal{N}(0,\Sigma_{0,n+1})$ models the observation noise. Then we can use these observations to provide a correction to the predictions in Eq.~\eqref{eqn:prior_update} as:
\begin{equation}\label{eqn:posterior_update}
    \begin{aligned}
        \bar{V}_{n+1|n+1}&=\bar{V}_{n+1|n}+K_{n+1}(d_{n+1}-G_{n+1}\bar{V}_{n+1|n}),\\
        R_{n+1|n+1}&=(\mathbb{I}-K_{n+1}G_{n+1})R_{n+1|n}\,,
    \end{aligned}
\end{equation}
where $\bar{V}_{n+1|n+1}$ and $R_{n+1|n+1}$ are called the posterior mean and covariance respectively. $K_{n+1}$ is called the Kalman gain matrix given by
\begin{equation}\label{eqn:Kalman_gain}
    K_{n+1}=R_{n+1|n}G_{n+1}^\top(G_{n+1}R_{n+1|n}G_{n+1}^\top +\Sigma_{0,n+1})^{-1}\,.
\end{equation}
{This procedure of updating mean and covariance $(V_{n+1|n+1},R_{n+1|n+1})$ from $(V_{n|n}, R_{n|n})$ using~\eqref{eqn:prior_update} and~\eqref{eqn:posterior_update} is called the Kalman filter.


Note a connection with section~\ref{sec:math_setup}, which also explains the derivation of Eqs.~\eqref{eqn:posterior_update}--\eqref{eqn:Kalman_gain}:
the Kalman filter equations in 
Eq.~\eqref{eqn:posterior_update}
and Eq.~\eqref{eqn:Kalman_gain} can be obtained as the closed-form solution to the problem of minimizing the cost function in Eq.~\eqref{eqn:cost_Bayesian_optimization}. However, note that this connection is valid when the evolution equations and observation equations are linear, and when the model error and observation noise are assumed to follow Gaussian statistics. 

A modified setup is needed for the purposes here, since
the measurements depend nonlinearly on the refractive index.
If the non-linearities are weak, we can approximate the measurement operator by linearizing about the prior mean and still use the Kalman filter updates that are described above. This is the idea behind the extended Kalman filter strategy, as follows. 

Let the observations now be given by the non-linear mapping:
\begin{equation}\label{eqn:d_filter}
   d_{n+1}= g_{n+1}(V_{n+1}) + \sigma_{n+1}\,,
\end{equation}
where $g_{n+1}$ is a non-linear operator and $\sigma_{n+1}\sim\mathcal{N}(0,\Sigma_{0,n+1})$ represents a Gaussian distributed observational error. Now we linearize $g_{n+1}$ around the prior mean as:
\begin{equation}\label{eqn:d_filter_linearized}
    g_{n+1}(V_{n+1})\approx g(\bar{V}_{n+1|n}) + G_{n+1}(V_{n+1}-\bar{V}_{n+1|n})\,,
\end{equation}
{where $G_{n+1}=\nabla_Vg_{n+1}|_{\bar{V}_{n+1|n}}$ is the Fr\'echet derivative of $g_{n+1}$ evaluated at $\bar{V}_{n+1|n}$. 
Let the model evolution and prior update still be given by Eq.~\eqref{eqn:V_model_eqn} and Eq.~\eqref{eqn:prior_update} respectively. 
Drawing the relation with Eq.~\eqref{eqn:data_KF}, we have:}
\begin{equation}\label{eqn:posterior_update_EKF}
    \begin{aligned}
                   \bar{V}_{n+1|n}&=F_{n+1}\bar{V}_{n|n},\\
        R_{n+1|n}&=F_{n+1}R_{n|n}F_{n+1}^\top +\Sigma_{n+1}\\
        \bar{V}_{n+1|n+1}&= \bar{V}_{n+1|n} + K_{n+1}(d_{n+1}-g(\bar{V}_{n+1|n}))\\
        R_{n+1|n+1}&= (\mathbb{I}-K_{n+1}G_{n+1})R_{n+1|n}\\
                K_{n+1}&= R_{n+1|n}G_{n+1}^\top(G_{n+1}R_{n+1|n}G_{n+1}^\top +\Sigma_{0,n+1})^{-1}\,.
    \end{aligned}
\end{equation}
This is the extended Kalman filter, in general form.
Note that it is the same as the (linear) Kalman filter
from Eq.~\eqref{eqn:prior_update}, 
Eq.~\eqref{eqn:posterior_update}
and Eq.~\eqref{eqn:Kalman_gain},
except the extended Kalman filter makes use of
the nonlinear observation operator $g$ and its linearization
from Eq.~\eqref{eqn:d_filter}--\eqref{eqn:d_filter_linearized}.

\subsection{Data assimilation using the PWE}
\label{subsec:data_assimi_PWE}
{To deploy the extended Kalman filter in the case of the PWE, we need to identify the model $F_{n+1}$ and the non-linear mapping $g_{n+1}$ of our interest and compute the Fr\'echet derivative $G_{n+1}$.}

An OU-process is used to describe the time evolution of the medium, as described in section~\ref{subsec:time-evol-medium}. At time $t_{n+1}$, the unknown state $V_{n+1}$ is represented by a vector of the Fourier coefficients $\hat{V}_{k_X,k_z}$ at time $t_{n+1}$ and its evolution is presented in the OU-process~\eqref{eqn:OU_process}. Plugging it into~\eqref{eqn:prior_update}, we have
\begin{equation}\label{eqn:prior_update_OU}
\begin{aligned}
        \bar{V}_{n+1|n}&= \bar{V}_{n|n}e^{-\gamma\Delta t},\\
        R_{n+1|n}&= R_{n|n}e^{-2\gamma\Delta t} + \frac{\sigma^2}{2\gamma}(1-e^{-2\gamma\Delta t})\,,
\end{aligned}
\end{equation}
where $\Delta t$ is the time lapse between observational measurements.
Note that a separate equation in the form of (\ref{eqn:prior_update_OU}) is needed for each wavenumber $(k_X,k_z)$ of the turbulent medium, so that
the quantities $V$, $\gamma$ and $\sigma$ in (\ref{eqn:prior_update_OU}) are meant to represent the quantities $\hat{V}_{k_X,k_z}$, $\gamma_{k_X,k_z}$ and $\sigma_{k_X,k_z}$, respectively, where the subscripts $(k_X,k_z)$ were suppressed to ease notation.

To specify the generic nonlinear observation map $g_{n+1}$, we deploy the PWE from \eqref{eqn:PWE} and take $g_{n+1}$ to be the measurement map $\mathcal{M}$ as defined in~\eqref{eqn:forward_operator}.
As a consequence, computing $G_{n+1}=\nabla_Vg_{n+1}$ calls for evaluating $\nabla_V\mathcal{M}_{j_1,j_2}$. Furthermore, if $V$ is parametrized by $M$ parameters $\{V_1,V_2,\cdots,V_M\}$, then the derivative $\partial_{V_m} \mathcal{M}_{j_1,j_2}$ is sought. These derivatives can be explicitly computed using the standard adjoint method, a technique from the calculus of variations. 
The details of the derivation are shown in the Appendix.
The result is
\begin{equation}\label{eqn:grad-M}
    \partial_{V_m} \mathcal{M}_{j_1,j_2}(V)=\int\limits_{z=0}^Z\int\limits_{X\in\mathbb{R}^2} \mathrm{Re}\Big\{ik A_{V,\phi_{j_1}}(X,z) h_{\psi_{j_2}}(X,z)\Big\}\frac{\partial V}{\partial V_m}\mathrm{d}X\mathrm{d}z\, , \quad m = 1,\cdots, M\,,
\end{equation}
where $h_{V,\psi_{j_2}}$ solves the adjoint equation using $A_{V,\phi_{j_1}}$ as the final data, i.e.,
\begin{equation}\label{eqn:adjoint}
\nabla_{X}^2h_{V,\psi_{j_2}}-2ik\partial_zh_{V,\psi_{j_2}}+k^2V h_{V,\psi_{j_2}}=0\,\ \text{with} \ h_{V,\psi_{j_2}}(X,z=Z)=A_{V,\phi_{j_1}}^\ast(X,Z)\psi_{j_2}(X)\,.
\end{equation}
This completes the specification of the data assimilation
methodology for the PWE.

\section{Results of computational experiments}\label{sec:numerical_exp}

In this section, we present the results of some computational experiments using the extended Kalman filter strategy. For numerical simulations, we consider a two-dimensional setting with $X=x$ with a source reference wave number of $k=2\pi\times 10^6$(rad/m). We assume that the receiver is placed at a distance of $Z=1500$ m away from the transmitter. There is a single source $\phi$ (hence $J_1=1$) taken as:
\begin{equation}\label{eqn:Gaussian_source}
    \phi(x)=\exp\Big(-\frac{x^2}{2r_s^2}\Big),\quad x\in[-a,a]\,,
\end{equation}
with source width $r_s=0.02$m and a transmitter window of half-width $a=0.05$m. The entire simulation is conducted over a much larger $x$-domain of $[-L,L]$ with half-length parameter $L=0.4$m. For simulating the forward model using the PWE, we use an operator splitting scheme as in~\cite{bao2002time}.
For the $x$-discretization, we use $\Delta x=0.002$ m and for $z$-discretization, we use $\Delta z = Z/256$ so that there are $256$ grid points along the $z$-direction. To model the medium, we consider an ansatz of the form
\begin{equation}\label{eqn:V_ansatz}
    V(x,z,t)=\epsilon (V_1(t)\sin(k_x x)+V_2(t)\cos(k_x x))\sin(k_z z)
\end{equation}
with $V_1, V_2$ being the spatial Fourier coefficients whose time evolution follows the OU processes:
\begin{equation}\label{eqn:OU_process_simul}
    \dot{V}_j=-\gamma V_j  + \sigma \dot{W}, \quad j=1,2\,.
\end{equation}
Here, $\gamma$ is chosen to be $1$s$^{-1}$. $\dot{W}$ denotes the standard Brownian motion and the noise level $\sigma$ is chosen as $0.1$. For simulating the OU process as a stochastic differential equation, we use an Euler-Maruyama scheme~\cite{platen1999introduction} with time step $\Delta t= 2^{-7}$s over a total simulated time of $T=1$s. 
For wavenumber parameters, $k_z$ is fixed at $k_z = 10\pi/Z$, and $k_x$ is set at different values: $k_x=m\pi/L$, with $m=2, 4, 6$ and $8$. As the strength of fluctuations, $\epsilon$ is taken as ${5}/{\sqrt{2}}\times 10^{-8}$. As observations, we consider intensity measurements of the form:
\begin{equation*}
    \int\limits_{x}|A|(x,Z)|^2\psi_{j_2}(x)\mathrm{d}x, \quad j_2=1,\cdots,J_2,
\end{equation*}
where the window functions $\psi_{j_2}$ have a Gaussian form over a receiver region $[-R,R]$ given by:
\begin{equation}
    \psi_{j_2}(x)=\frac{1}{\sqrt{2\pi}r_w}\exp\Big(-\frac{(x-x_{j_2})^2}{r_w^2}\Big),\quad x\in[-R,R]\,,
\end{equation}
where $x_{j_2}$ denotes the centers of the receiver windows. In the simulation, we choose $J_2=50$, with receiver region of width $R=0.05$m and each receiver window function with half-width $r_w =0.004$m.

As an initial view of the problem setting, before considering
the reconstruction of the medium using data assimilation,
we consider the cost function landscape and the intensity measurements.
In Fig.~\ref{fig:cost_fn}, we plot the cost functions from the measurement operators given by:
\begin{equation}\label{eqn:cost_fn_expt}
    \mathcal{L}(V)=\frac{1}{J_2}\sum_{j_2=1}^{J_2}|\mathcal{M}_{j_2}(V)-d_{j_2}|^2\,,
\end{equation}
with $V=(V_1,V_2)$ and the data $d_{j_2}$ is generated by running PWE with the ground truth media. It can be observed that the cost function landscapes become more convex as the frequency $k_x$ increases, which suggests the recovery may be more successful in those cases.
\begin{figure}
    \centering
    \includegraphics[width=14cm]{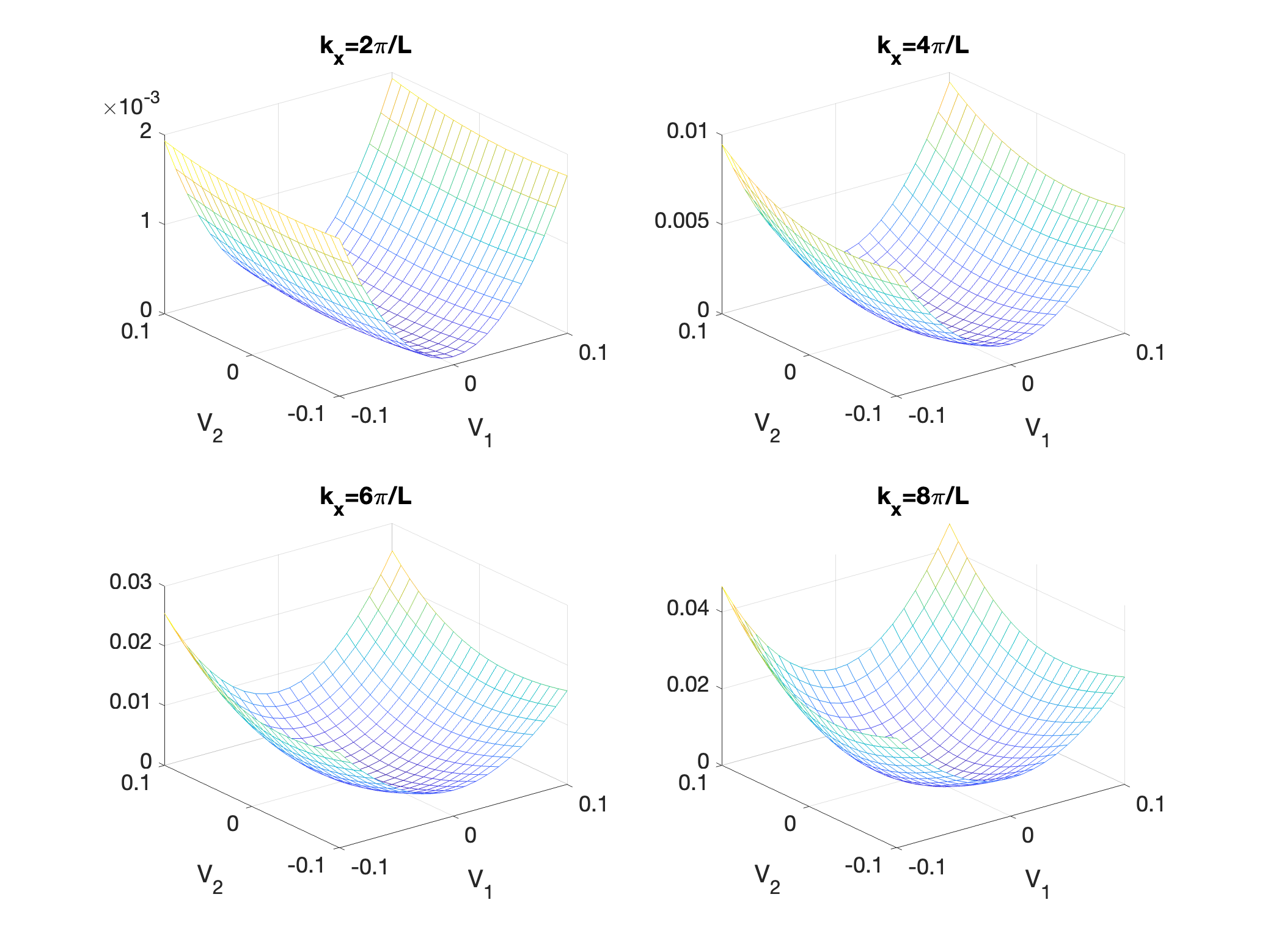}
    \caption{The behavior of the cost function~\eqref{eqn:cost_fn_expt} with $V$ specified in~\eqref{eqn:V_ansatz}.}
    \label{fig:cost_fn}
\end{figure}
In Fig.~\ref{fig:OU_process}, we plot a realization from simulating the OU-process in Eq.~\eqref{eqn:OU_process_simul},
which describes the Fourier coefficients $V_1(t)$ and $V_2(t)$ of the medium and their evolution in time. 
The corresponding intensity measurements over the receiver region are plotted in Fig.~\ref{fig:observation} during the entire duration of the simulations. The noise level in the observations $\Sigma_{0,n}$ for all $n$ is taken as $2.5\times 10^{-3}\mathbb{I}$. The effect of the turbulence becomes more prominent as the frequency $k_x$ increases. 

Now we present the results of the reconstruction algorithm using data assimilation.
In Fig.~\ref{fig:V1_recons} and Fig.~\ref{fig:V2_recons}, we plot the reconstruction of the Fourier coefficients $V_1$ and $V_2$ as a function of time. The tracking is seen to be accurate for larger frequencies. Finally in Fig.~\ref{fig:source_recov_real} and Fig.~\ref{fig:source_recov_abs}, we plot the effect that parameter estimation has on source recovery. In the plot, the true source is set to be a Gaussian of the form Eq.~\eqref{eqn:Gaussian_source}. The reconstruction of the source using a guess of $V=0$ is presented as a comparison. As expected, in cases where the parameter estimation was accurate, the reconstruction of the source is more accurate as well.
\begin{figure}
    \centering
    \includegraphics[width=7cm]{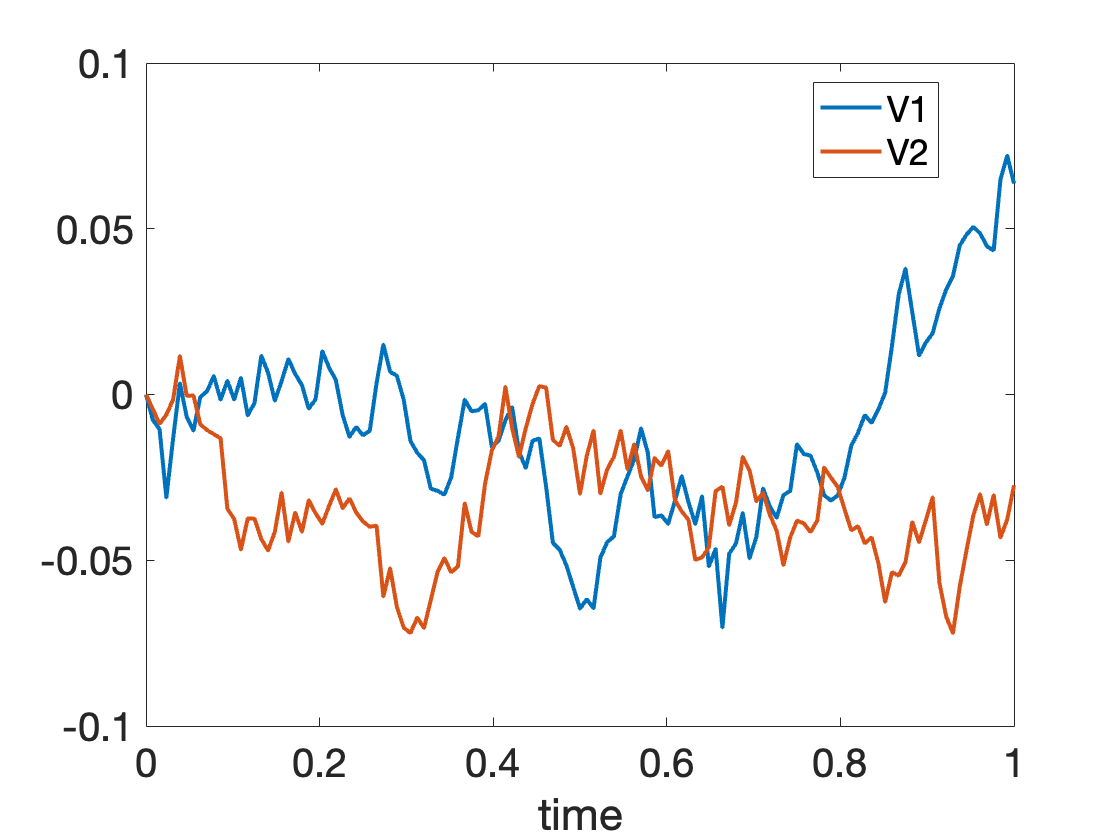}
    \caption{ Time series of Fourier coefficients of the medium, $V_1(t)$ and $V_2(t)$, generated by simulating the OU processes Eq.~\eqref{eqn:OU_process_simul} with parameters $\gamma=1$ s$^{-1}$ and $\sigma = 0.1$. Time is in seconds.}
    \label{fig:OU_process}
\end{figure}

   

 \begin{figure}
    \centering
  \includegraphics[width=13cm]{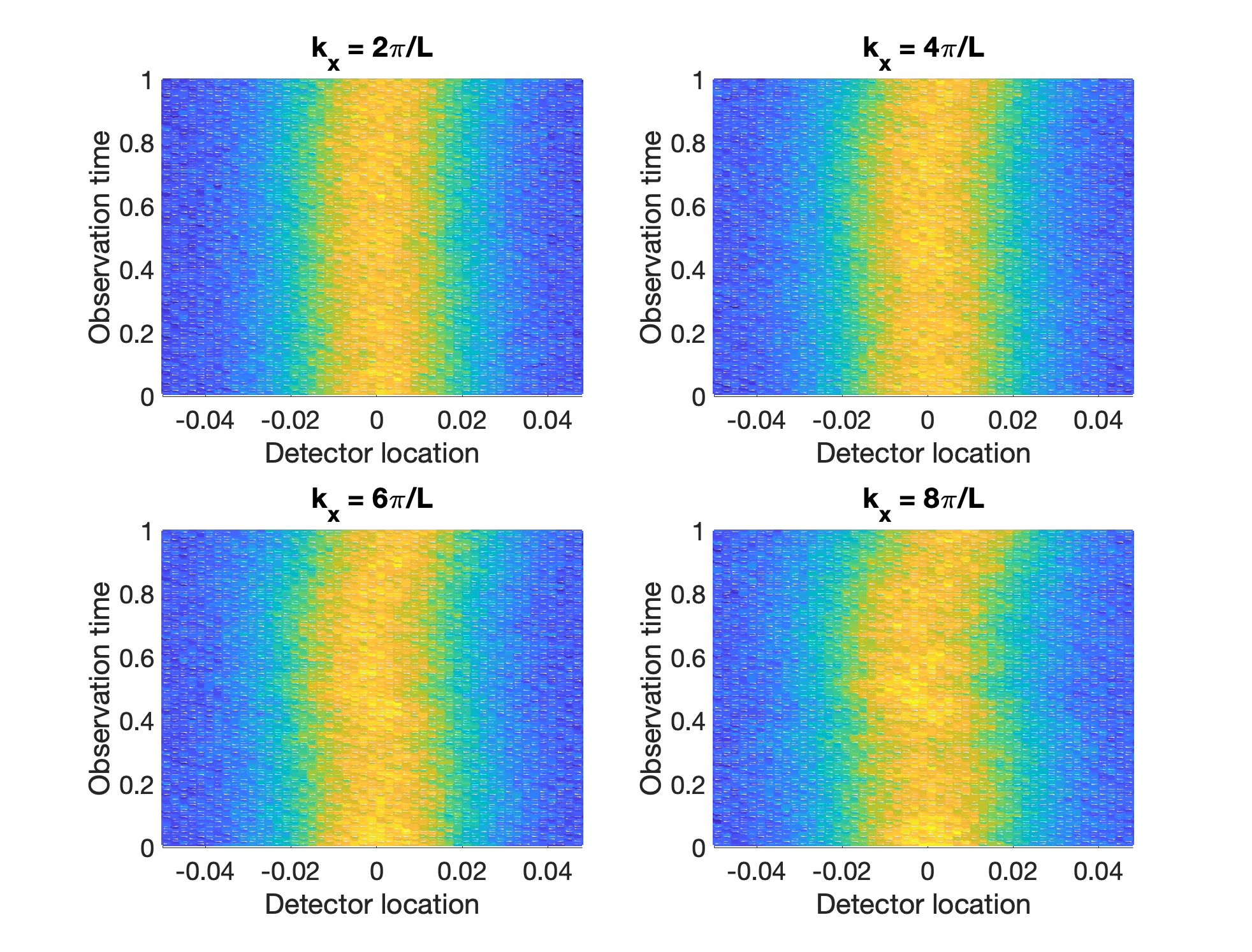}
    \caption{ Intensity measurements recorded at the receiver region.}
     \label{fig:observation}   
    \end{figure}

    \begin{figure}
   \begin{subfigure}{0.99\textwidth}   
    \centering
       \includegraphics[width=11cm]{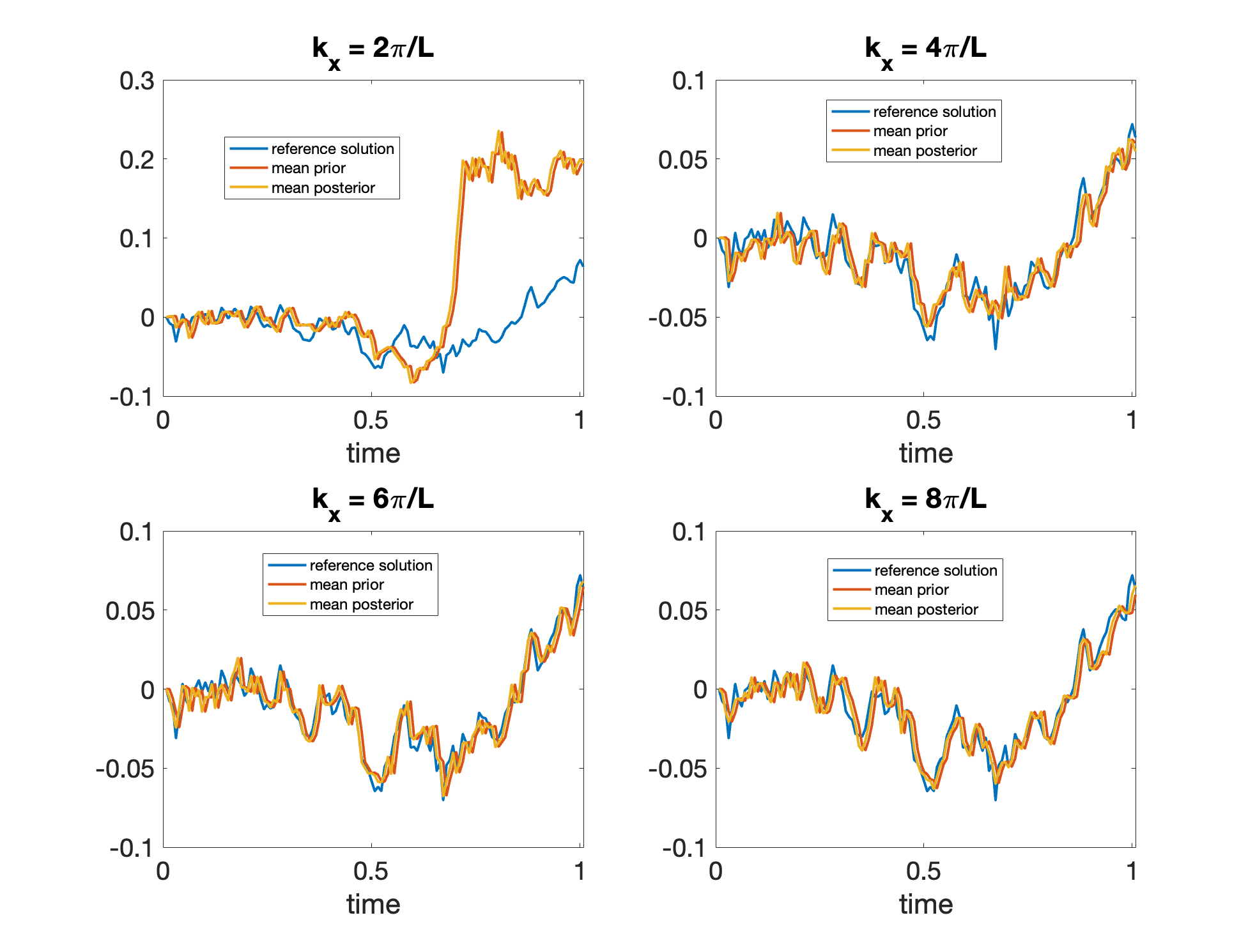}
    \caption{ $V_1$}
    \label{fig:V1_recons}
\end{subfigure}
   \begin{subfigure}{0.99\textwidth}   
    \centering
  \includegraphics[width=11cm]{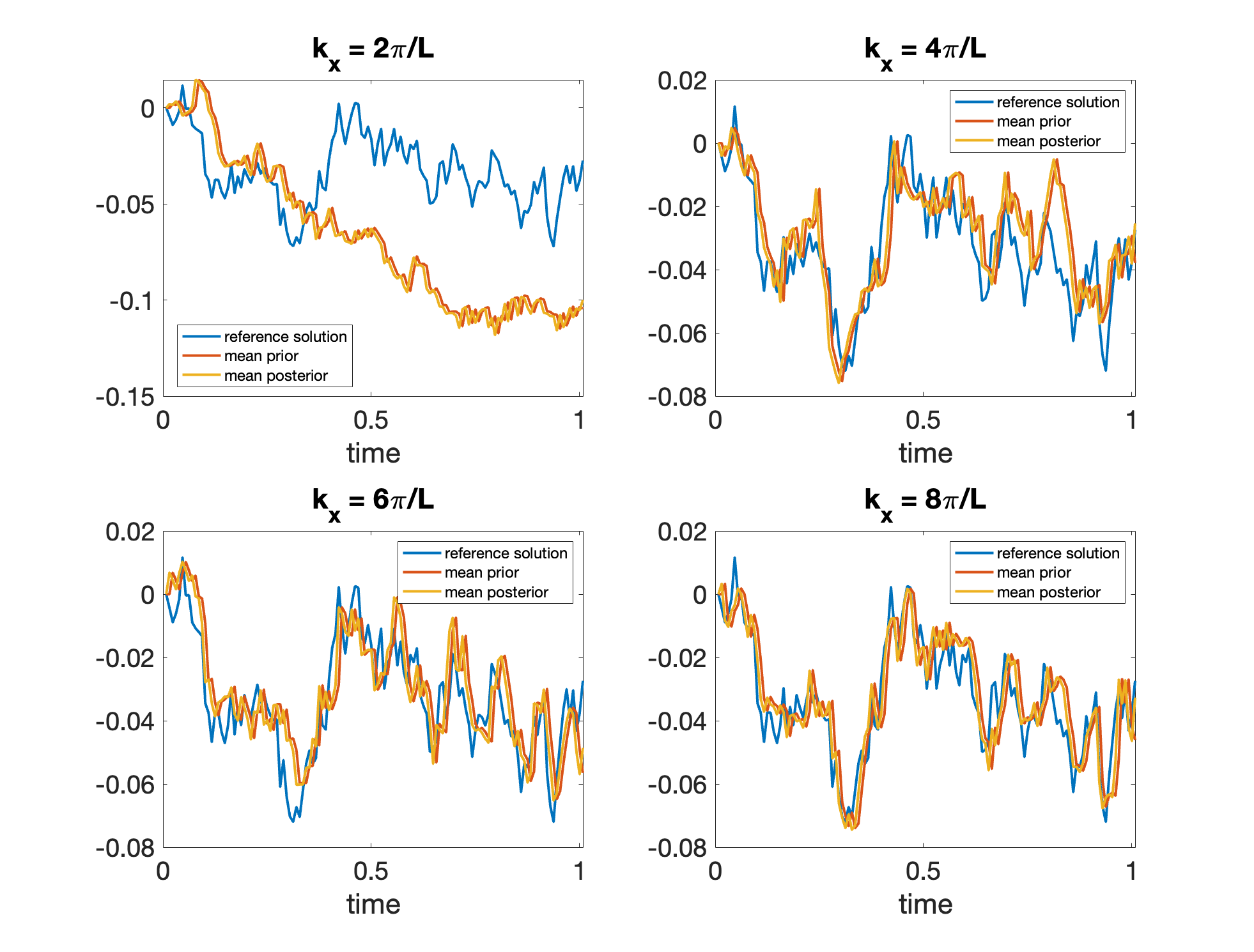}
    \caption{ $V_2$}
    \label{fig:V2_recons}
    \end{subfigure}
    \caption{Recovery of the Fourier coefficients $V_1(t)$ and $V_2(t)$ of the refractive index, from (\ref{eqn:V_ansatz}), and their evolution over time as measurements are received and assimilated. The Fourier coefficients $V_1$ and $V_2$ characterize the spatial variations of the refractivity. The recovery is successful when the wavenumber $k_x$ is larger ($4\pi/L, 6\pi/L, 8\pi/L$), and it is inaccurate when $k_x$ is smaller ($2\pi/L$) at the later times of the experiment. }
    \end{figure}

        \begin{figure}
   \begin{subfigure}{0.99\textwidth}   
    \centering
       \includegraphics[width=13cm]{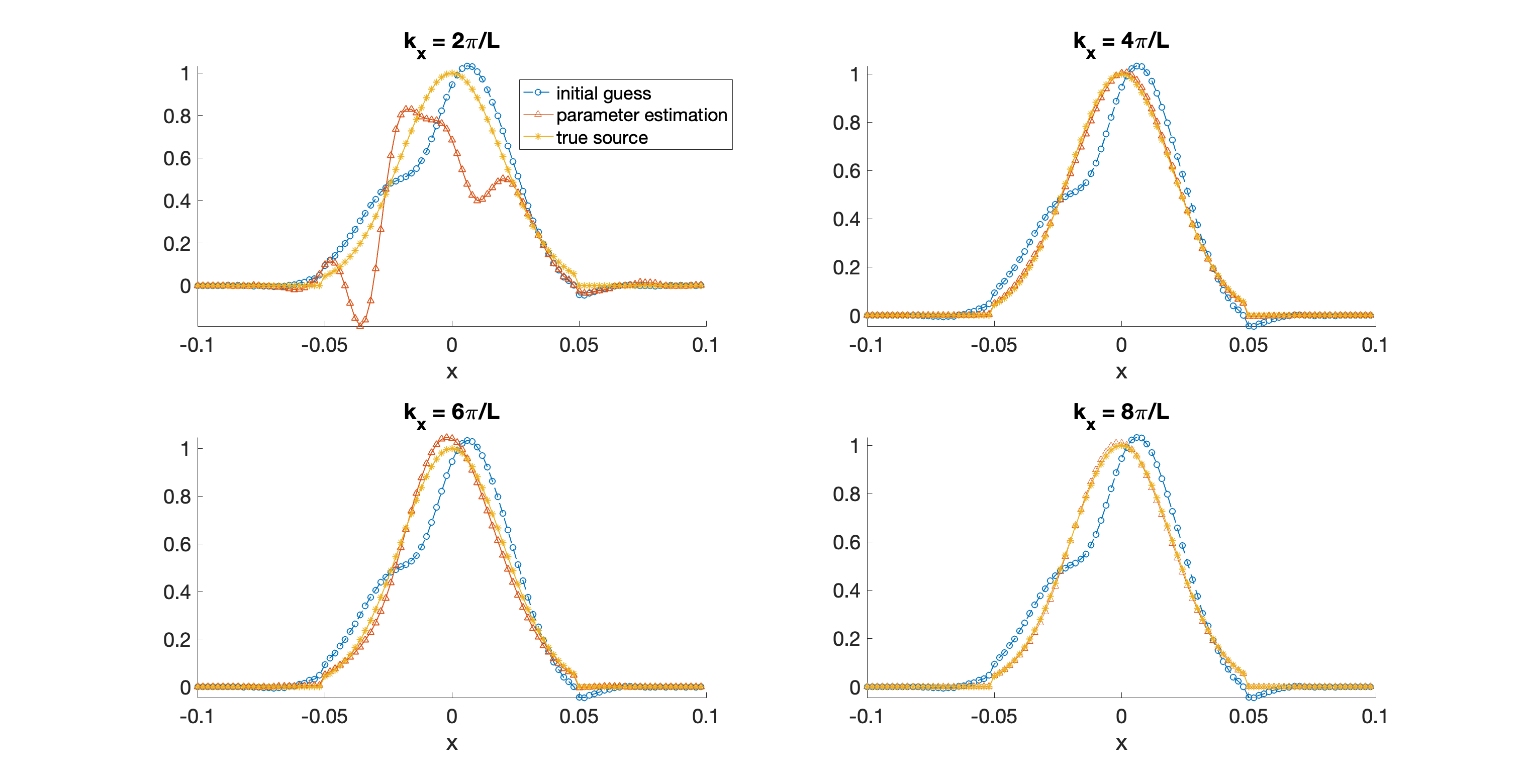}
    \caption{ Real part}
    \label{fig:source_recov_real}
\end{subfigure}
   \begin{subfigure}{0.99\textwidth}   
    \centering
  \includegraphics[width=13cm]{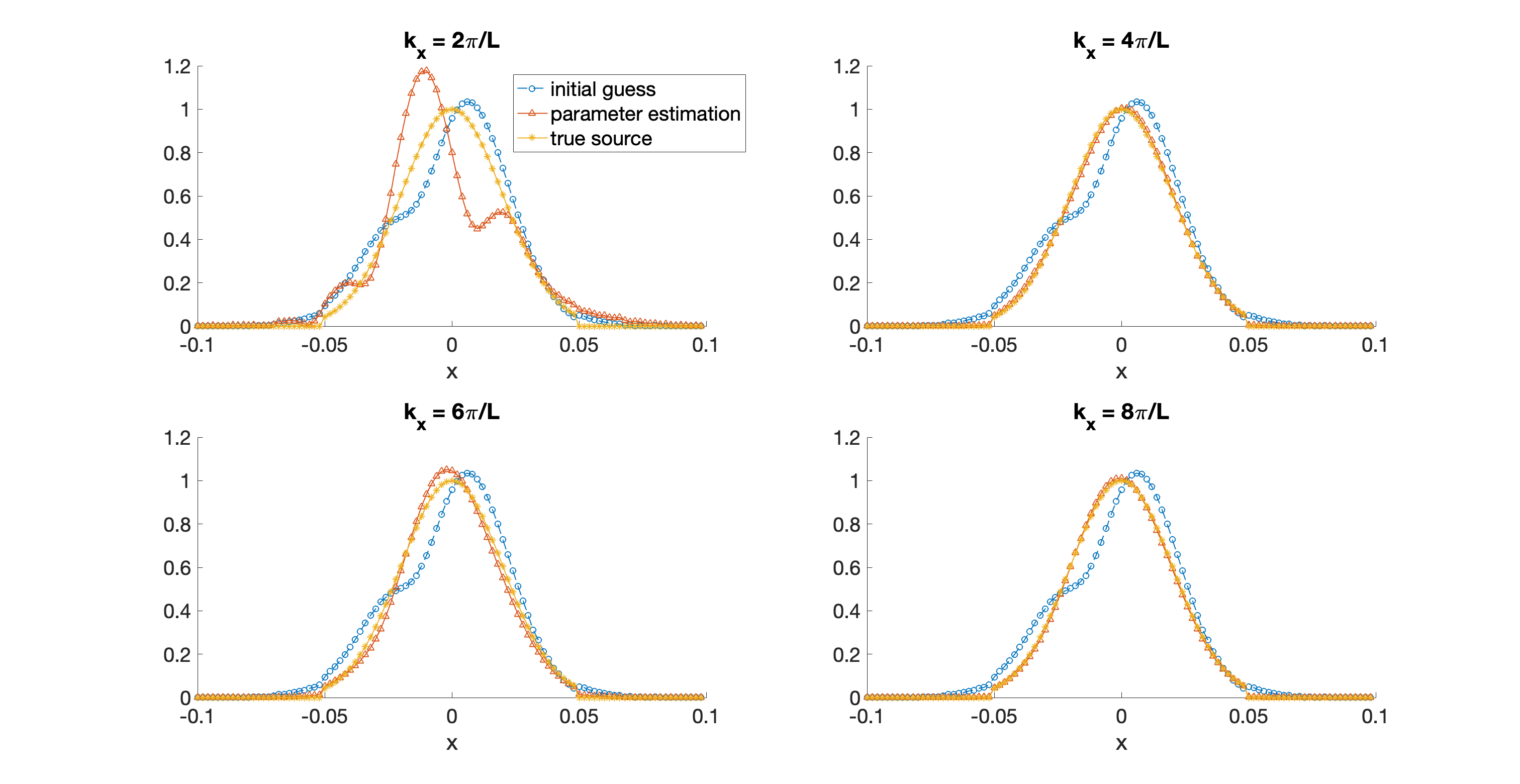}
    \caption{ Absolute value}
    \label{fig:source_recov_abs}
    \end{subfigure}
    \caption{Reconstruction of the source using the reconstructed parameters $V_1 $ and $V_2$. The profile of the true source and the reconstructed source using an initial guess of $V_1=V_2=0$ are shown in comparison. The reconstruction is more accurate when the wavenumber $k_x$ is larger ($4\pi/L, 6\pi/L, 8\pi/L$) and less accurate when $k_x$ is smaller ($2\pi/L$).}
    \end{figure}

\clearpage
\section{Concluding discussion}

In this paper, our goal was to investigate the potential
applicability of data assimilation methods in the setting
of optical communications. This approach is partly motivated
by the success of data assimilation and filtering methods
in many other areas of science and engineering.

In particular, the aim here was to infer information
about the spatial and temporal variations of the
turbulent medium through which the optical signal
is propagating.
Given measurements of beam intensity at a receiver,
the computational results suggest that the state of the
medium can be recovered successfully in many cases,
while in other cases the errors were larger.
These computational results are promising, and they provide motivation for further investigation in the future.

\section*{Appendix}\label{sec:Appendix}
As discussed in Section~\ref{sec:math_setup}, Eq~\eqref{eqn:forward_operator} defines a map $\mathcal{M}_V$ that returns a value when the source $\phi$ and measurement window function $\psi$ is specified. We now provide the computation of the derivative on $V$ of $\mathcal{M}_V$. This derivative is needed for the extended Kalman filter algorithm in (\ref{eqn:grad-M}). Throughout the section, since the context is clear, for notational convenience, we omit $(\phi,\psi)$ dependence.

Taking the Fr\'echet derivative of a functional on a function is a standard practice in the field of calculus of variations. A fixed set of mathematical strategy that deploys the adjoint method can be utilized. We summarize the finding in the following theorem.
\begin{theorem}
    Denote by $A$ the solution to the PWE~\eqref{eqn:PWE} with medium being the function $V$ and the source being $\phi$, and assume the receiver is presented by the window function $\psi$, then for this fixed configuration of source and receiver $(\phi,\psi)$, the Fr\'echet derivative of the forward operator $\mathcal{M}$ in Eq.~\eqref{eqn:forward_operator} is given by:
\begin{equation}\label{eqn:grad}
    \nabla_V \mathcal{M}(V) = \mathrm{Re}\Big\{ ik A(X,z) h(X,z)\Big\}\,,
\end{equation}
where $h$ solves the adjoint equation
\begin{equation}\label{eqn:adjoint}
    \begin{aligned}
                    \nabla_{X}^2h-2ik\partial_zh+k^2V h&=0,\\
                    h(X,z=Z)&=A^\ast(X,Z)\psi(X)\,.
    \end{aligned}
\end{equation}
\end{theorem}
\textit{Proof:} Let $A$ solve~\eqref{eqn:PWE} using the medium $V$ with the source $\phi$, and let $A_1$ be the solution to~\eqref{eqn:PWE} with the perturbed medium $V+\delta V$. Then up to $\mathcal{O}(\delta V)$, their difference $\Tilde{A}=A_1-A$ satisfies the following partial differential equation:\begin{equation}
\begin{aligned}
        \nabla_{X}^2\Tilde{A}+2ik\partial_z\Tilde{A}+k^2\delta VA+k^2V\Tilde{A}&=0,\\
         \Tilde{A}(z=0,X)&=0\,.
\end{aligned}
\end{equation}
Let $h$ denote the solution to the adjoint equation Eq.~\eqref{eqn:adjoint}. Multiplying the equation for $\Tilde{A}$ with $h$ and vice-versa, integrating with respect to $X\in\mathbb{R}^2,z\in(0,Z)$ and subtracting the resultant equations gives
\begin{equation}
    \int\limits_{z=0}^Z \int\limits_{X\in\mathbb{R}^2}  [h\nabla_{X}^2\Tilde{A}-\Tilde{A}\nabla_{X}^2h+2ikh\partial_z\Tilde{A}+2ik\Tilde{A}\partial_z h+k^2\delta V A h]\mathrm{d}X\mathrm{d}z=0\,.
\end{equation}
Using the boundary condition that $\Tilde{A}(X,z=0)=0$, we have
\begin{equation}\label{eqn:adj_perturb}
    \begin{aligned}
\int\limits_{X\in\mathbb{R}^2} h(X,Z)\Tilde{A}(X,Z)\mathrm{d}X&= \frac{ik}{2}\int\limits_{z=0}^Z \int\limits_{X\in\mathbb{R}^2} \delta V A h\mathrm{d}X\mathrm{d}z\,.
    \end{aligned}
\end{equation}
Using the boundary information of $h$ at $z=0$ and noticing the left hand side of Eq.~\eqref{eqn:adj_perturb} is the perturbation of measurement $\mathcal{M}(V+\delta V)-\mathcal{M}(V)$. Hence we have derived that the derivative of $\mathcal{M}$ over $V(X,z)$ is Eq.~\eqref{eqn:grad}.

In addition, if the function $V(X,z)$ is parametrized by $M$ parameters $\{V_1,V_2,\cdots,V_M\}$ (e.g., the parameters could be the Fourier coefficients), the derivative of $\mathcal{M}$ with respect to each parameter, according to the chain rule, is given by:
\begin{equation}
    \partial_{V_m} \mathcal{M}(V)=\int\limits_{z=0}^Z\int\limits_{X\in\mathbb{R}^2} \mathrm{Re}\Big\{ik A(X,z) h(X,z)\Big\}\frac{\partial V}{\partial V_m}\mathrm{d}X\mathrm{d}z\, \quad m = 1,\cdots, M\,.
\end{equation}

\section*{Backmatter}

\begin{backmatter}
\bmsection{Funding}
The research of Q.L. is partially supported by Office of Naval Research (ONR) grant N00014-21-1-2140, and the research of A.N. and S.N.S. is partially supported by ONR grant N00014-21-1-2119.


\bmsection{Disclosures} The authors declare no conflicts of interest.

\bmsection{Data Availability Statement} 
Data and code underlying the results presented in this paper are available in~\cite{Nair2024_PWE_code}.
\end{backmatter}


\bibliography{Reference}






\end{document}